# A Collaborative Approach to Angel and Venture Capital Investment Recommendations

Xinyi Liu[*] and Artit Wangperawong

[*] New York University, New York, NY, USA.
To whom correspondence should be addressed; Email: xl2351@nyu.edu, xinyiliu0227@gmail.com

**Abstract**

Matrix factorization was used to generate investment recommendations for investors. An iterative conjugate gradient method was used to optimize the regularized squared-error loss function. The number of latent factors, number of iterations, and regularization values were explored. Overfitting can be addressed by either early stopping or regularization parameter tuning. The model achieved the highest average prediction accuracy of 13.3%. With a similar model, the same dataset was used to generate investor recommendations for companies undergoing fundraising, which achieved highest prediction accuracy of 11.1%.

## 1. Introduction

Since detailed information about private companies, especially startups, tend to be unavailable to the public, angel and venture capital (VC) investors would have to spend considerable time and resources to discover new companies to invest in. For startup valuations, researchers have gone through great lengths to conduct questionnaires and interviews to obtain the necessary data [1]. On the other hand, studies also found that financial information does not play a major role for how VC firms choose investments [2], as early stage ventures typically cannot provide mature financial data necessary for traditional valuation techniques [3], which makes it even more difficult for angel and VC investors to narrow down the potential options to consider.

Many times investors rely on their relationships with a company's incumbent investors or a funding round's lead investor, thereby limiting their options for investing [4]. Investors may also more successfully invest in companies for which they can leverage their networks and resources [5]. Successful investment leads are therefore not necessarily the most demanded startups because an investor might not have the opportunity to engage in such a deal. Here we present a way to filter potential investment options for investors to investigate, so that they can spend time and resources more efficiently on companies that they have higher potential to close a deal on.

### 1.1 Application

An investment individual or firm tends to have an investment thesis, which dictates the scope of their investment activities [6]. For example, investors that focus on a few specific industries or geography would likely prefer the same companies [7]. Investors might also invest in companies that can hedge the risk of their investment portfolio [8]. Even if there were not lucid reasons given for investing in a company, one investor's decision whether to invest or not can still provide positive or negative signals to affect other investors' decisions [9]. Hence, we developed a model without relying on detailed financial analysis to recommend potential investment targets for a given investor based what other similarly behaving investors have invested in. Such an approach may capture the investor's direct or indirect investing network.

### 1.2 Collaborative Filtering

Collaborative filtering is a commonly used technique involving machine learning algorithms to generate movie recommendations based on the viewing preferences of users in the system [10]. Among the sets of collaborative filtering algorithms, here we used Matrix Factorization (MF), or Matrix Decomposition, based on the alternating least squares (ALS) optimization method. A matrix of user ratings comprising of rows and columns representing users and movies, respectively, can be factored into a product of two matrices in order to predict the unknown values. In our case, we factored the matrix $M$ that represents whether an investor $i$ invested in a company $c$, where elements of matrix $M$ are:

$$m_{ci} = \begin{cases} 1 \text{ if } i \text{ invested in } c \\ 0 \text{ if } i \text{ did not invest in } c \end{cases} \quad \textbf{(Eqn 1)}$$

into a product of a factor matrix that represents investors and a factor matrix that represents companies. For the factor matrix $X$ representing companies, each row $x_c$ contains the latent factors about a company, while for the factor matrix $Y$ representing investors, each row $y_i$ contains the latent factors about an investor.

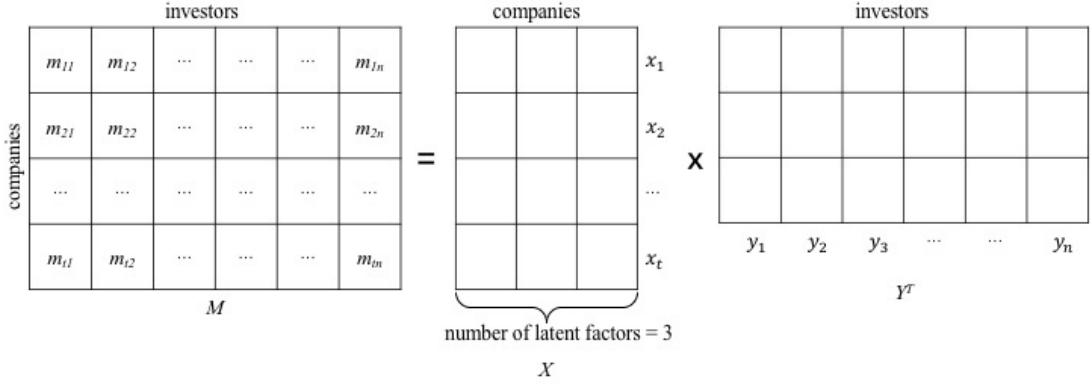

**Figure 1.** Matrix factorization diagram for investors and companies. For illustrative purposes, the rows $y_i$ of matrix $Y$ are shown as the columns of $Y^T$.

## 2. Methodology

Here we discuss how we applied collaborative filtering to the dataset from Crunchbase in 2013 [11]. The dataset contains 21,417 companies and 16,946 investors making a total of 80,245 investments, including repetitive investments by an investor in the same company. In our model, we found that accounting for multiple investments in the same company by the same investor does not improve accuracy, which might be due to the practice of follow-on investments by venture capitals obligated to mitigate signaling issues [12]. Therefore, we only used unique investor-company pairs, dropping duplicates from the dataset.

In addition to recommending companies to investors, we can use the same dataset to recommend investors to companies by transposing the matrix $M$ and factoring it into the product of a matrix that represents investors and a matrix that represents companies.

### 2.1 Loss Function

The loss function used for ALS is

$$L = \sum_{c,i}(m_{ci} - x_c^T y_i)^2 + \lambda \left( \sum_c \|x_c\|^2 + \sum_i \|y_i\|^2 \right)$$

**(Eqn 2)**

where the first part of this function is the sum of squared errors, and the second part is the regularization term to address overfitting [13].

Each iteration is based on alternatingly minimizing the loss function above for $X$ and $Y$. First, $X$ and $Y$ are randomly initialized according to the number of investors, companies and latent factors. $X$ is fixed while updating $Y$ using three steps of the conjugate gradient method, then $Y$ is fixed while updating $X$ using three steps of the conjugate gradient method. We can consider that the algorithm converges when the change to $X$ and $Y$ is negligible with subsequent iterations. Finally, we can rank our recommendations by the values $\hat{m}_{ci}$ of the prediction matrix $\hat{M}$ that correspond to $m_{ci} = 0$ in the input matrix $M$. The value in a given entry specified by $\hat{m}_{ci}$ reflects the strength of the recommendation for investor $i$ to invest in company $c$.

### 2.2 Accuracy

While many prominent studies involving matrix factorization (e.g., Netflix prize competitors) compare performance based on MAE or RMSE [14-16], here we used test accuracy as a more relevant indicator of the effectiveness of our models with MSE, namely training loss, as a complement. Our goal is to maximize the predictive power of performing binary classification for each investor-company pair.

To calculate the accuracy, we randomly hid one investment from the set of investments of a random sample of 10% of investors that have invested twice or more. These hidden investments can be considered as test data. We wanted to ensure that each investor had adequate investment history for us to generate recommendations for. A prediction is considered to be correct if the hidden investment shows up in the top-10 generated investment recommendations for the investor. Therefore, the accuracy is calculated as:

$$Accuracy = \frac{Correct\ predictions}{Hidden\ investments} \quad \textbf{(Eqn 3)}$$

The prediction accuracy we report is the average of fifty trials.

### 2.3 Training

In order to determine the optimal number of iterations and latent factors, we first adjust the iterations to see the optimal number of iterations and then adjust the number of latent factors to gain the highest prediction accuracy. Note that while loss is reported below for the training dataset, the prediction accuracy is based on test dataset as explained above.

## 3. Results and Discussion

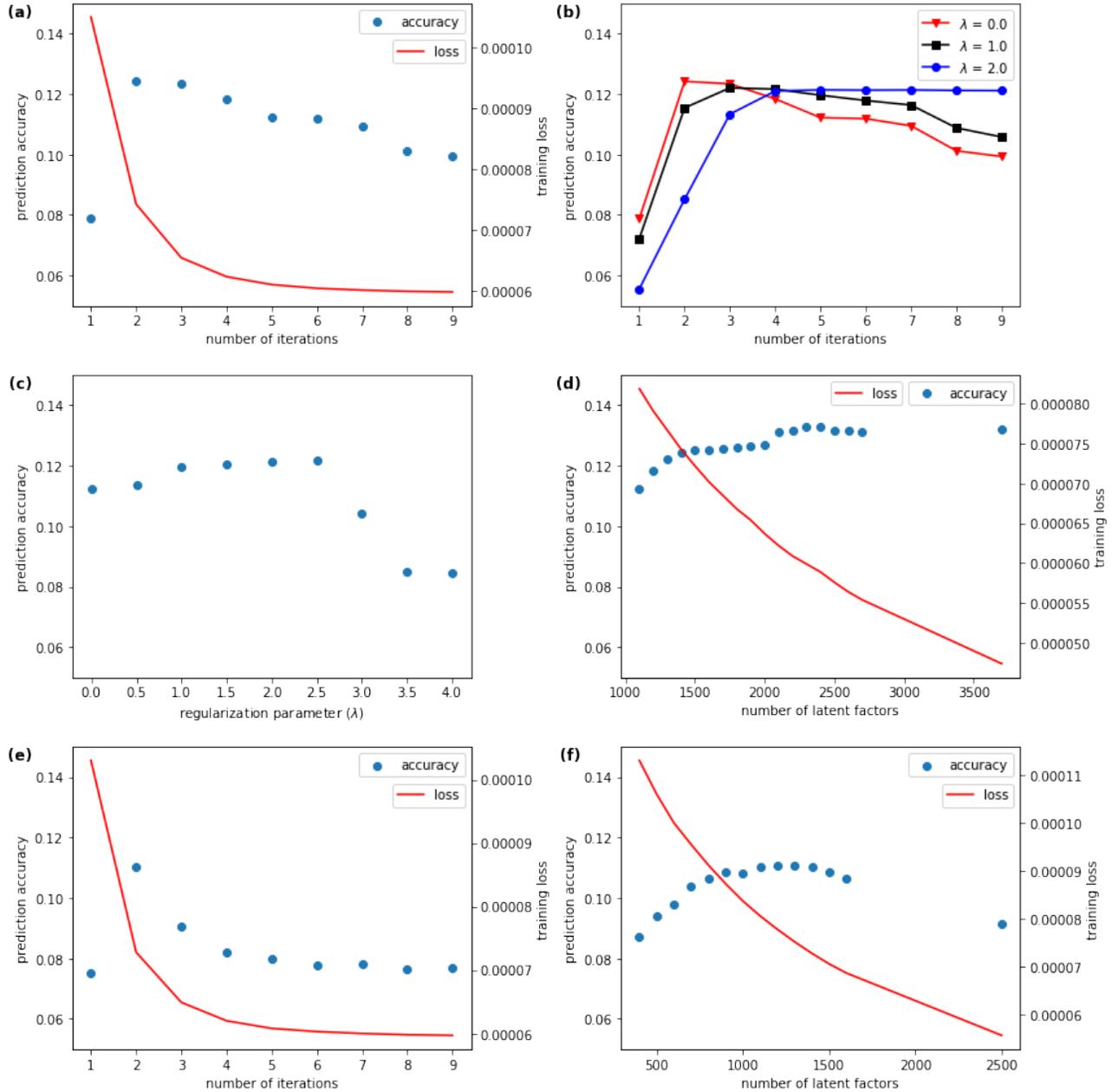

**Figure 2.** (a-d) For recommending companies to investors, and (e-f) for recommending investors to companies. (a) Effect of iterations on accuracy and training loss at 1400 latent factors and $\lambda = 0$. Beyond 2 iterations, accuracy does not increase any further although the training loss keeps decreasing. (b) Comparison of accuracies between $\lambda = 0$, $\lambda = 1$ and $\lambda = 2$ at 1400 latent factors. Beyond four iterations, increasing regularization improves accuracy more substantially with iterations. (c) Effect of regularization on accuracy at 1000 latent factors and 8 iterations. The highest accuracy for 1400 latent factors and 5 iterations, which is about 12.2%, was attained with $\lambda = 2.5$. (d) Effect of latent factors on accuracy at 2 iterations and $\lambda = 0$. At 2400 latent factors, the highest accuracy achieved was 13.3%. However, higher than 2200 latent factors produced comparable accuracies. (e) Effect of iterations on accuracy at 1400 latent factors and $\lambda = 0$. Beyond 2 iterations, accuracy does not increase any further while the training loss keeps decreasing. (f) Effect of latent factors on accuracy at 2 iterations. The highest accuracy, which is about 11.1%, was attained with 1300 latent factors.

On a 2.7 GHz Intel core i7 with 4 cores and 16 GB RAM, each iteration takes about 8 seconds at 1400 latent factors, and up to about 40 seconds at 2400 latent factors. Training was done in parallel on multiple cores.

As shown in Fig. 2(a), with increasing iterations, both accuracy and training loss decrease after 2 iterations, while the rate of training loss also decreases significantly. These results indicate that the model is overfitting after two iterations. While early stopping can be a solution to overfitting [17], in production deployment the peak at which to perform early stopping might vary and be easily missed. Regularization can therefore be a parameter to more reproducibly address overfitting, yielding higher test accuracies in practice. For example, Fig. 2(b) shows that for $\lambda = 1.0$ the accuracies are comparable between 3 and 6 iterations, thereby offering an operating range with some margin of stability.

In order to see if increasing the regularization parameter could possibly improve prediction accuracy at a higher number of iterations, we varied $\lambda$ at 5 iterations (Fig. 2(c)). It appears that $\lambda > 2.5$ only results in lower accuracies.

In spite of the ability to regularize the models, early stopping may still be a valid approach to reduce the computational cost for training. Therefore, in Fig. 2(d) we varied the number of latent factors to find the optimal number for the highest accuracy at two iterations. However, it should be noted that increasing the number of latent factors increases computational cost because of the larger matrices involved in the calculations. Therefore, a combination of early stopping and high number of latent factors may offer the best tradeoffs between accuracy and training computational cost.

Table 1 below lists a few examples of company recommendations for investors. Out of respect for the investor's privacy, we do not disclose investor names, although the information is available in the database.

**Table 1.** Examples of correct predictions for anonymized investors. In the second row, included are known investments prior to prediction.

| Investor A | Investor B | Investor C |
|---|---|---|
| CreditPing.com AwesomenessTV DramaFever | Square ChallengePost | Cozi Group Livestream 4INFO Topix Wanderful Media |
| Tubular | Fluidinfo | Ongo |

Our model recommends Tubular to Investor A, an individual angel investor investing in seed rounds. This recommendation makes sense since Investor A has demonstrated interest in the digital media industry through his investments, AwesomenessTV and DramaFever, both of which are video platforms. Tubular provides video intelligence software that refines content and marketing strategies for video platforms. Additionally, since Investor A is located in the Greater Los Angela Area, he appears to prefer investments based in California. CreditPing.com and AwesomenessTV are all based in California, while DramaFever has a California-based parent company, Warner Bros. Tubular is also in California. Moreover, AwesomenessTV is one of Tubular's customers, which suggests a relationship between AwesomenessTV and Tubular that Investor A might be able to leverage.

In the second example, our model recommends Fluidinfo to Investor B, another individual seed investor. In this case, the two known investments, Square and ChallengePost, are based in different industries (payment processing and software recruiting) and locations (San Francisco and New York). Therefore, Investor B did not nominally appear to have an industrial or geographic preference. A brief review of these three companies reveals that they all share another common investor besides Investor B. This indicates that an investor may follow other investors based on their network or other interests.

In the last example, Investor C is a media holding company. Based on its investments in Cozi Group, Livestream, 4INFO, Topix and Wanderful Media, Investor C appears to focus on internet media investment. 4INFO is a company that collects data and keeps track of the mobile advertisement for consumers. Cozi Group helps families organize their calendars and at the same time provides articles and recipes that targeting modern families. Topix began as an aggregation of news stories and now works on content creation. Therefore, it is unsurprising that the model recommends Ongo, a company that also provided content to users from trusted news brands as Investor C's other investments.

We also transposed the matrix $M$ and performed the same training process to obtain the accuracy of recommending investors for companies. As before, two iterations was found to be most suitable for this model (Fig. 2(e)). At two iterations, the effect of the number of latent factors on prediction accuracy varies differently than the prior model above. As shown in Fig. 2(f), increasing the number of latent factors beyond 1300 appeared to only reduce accuracy. This model may be useful for companies searching for investors when fundraising for their ventures.

As demonstrated, our model for recommending investments can capture various types of investing, e.g. geography, industry, connections, etc. It can also recommend investors to companies. Accuracy may be improved by adding more data to the training set. For example, we can further improve the training with investments that happened after 2013. It may also be useful to obtain data about the amount of each investor's investment to capture the strength of an investment and produce a model not solely based on binary values. Group investing, e.g. syndicates and funding rounds, may be an additional factor as simultaneously investing in the same company might be better considered as a different signal to the model.

## 4. Conclusion

We demonstrate in this work that collaborative filtering can be used to predict investments by angel and venture capital investors. Although we do not seek to quantify or support any returns on or profitability of such investments, the framework presented can be used to recommend relevant companies to investors and vice versa. Investment platforms such as Crunchbase [11] and AngelList [18] can serve more relevant content to their users. Investors and companies undergoing investing and fundraising can narrow down their prospects to save both time and costs.